\documentclass[%
 reprint,
 amsmath,amssymb,
 aps,
]{revtex4-2}

\usepackage{graphicx}
\usepackage{dcolumn}
\usepackage{bm}

\usepackage{booktabs}
\usepackage{dsfont}
\usepackage[dvipsnames]{xcolor}
\usepackage{placeins}

\newcommand{\nil}{{\color{white}(0)}}
\newcommand{\blank}{{\color{white}0}}

\newcommand{\pP}{$p\mathds{P}          \rightarrow t\bar{t}$}
\newcommand{\gamp}{$\gamma p             \rightarrow t\bar{t}$}
\newcommand{\PP}{$\mathds{P}\mathds{P} \rightarrow t\bar{t}$}
\newcommand{\gamgam}{$\gamma\gamma         \rightarrow t\bar{t}$}
\newcommand{\aP}{$\gamma\mathds{P}     \rightarrow t\bar{t}$}

\newcommand{\pPttbar}{$\sigma_{(p\mathds{P} \rightarrow t\bar{t})}$}
\newcommand{\apttbar}{$\sigma_{(\gamma p \rightarrow t\bar{t})}$}
\newcommand{\PPttbar}{$\sigma_{(\mathds{P}\mathds{P} \rightarrow t\bar{t})}$}
\newcommand{\aattbar}{$\sigma_{(\gamma\gamma \rightarrow t\bar{t})}$}
\newcommand{\aPttbar}{$\sigma_{(\gamma\mathds{P} \rightarrow t\bar{t})}$}

\begin{document}


\title{Elastic Potential: A proposal to discover elastic production of top quarks at the Large Hadron Collider}

\author{James Howarth}
\affiliation{Department of Physics and Astronomy, University of Glasgow.}

\date{\today}

\begin{abstract}
In this letter, I present the motivation and an example analysis method for discovering the elastic production of top-quark pairs at the LHC using forward proton tags, including an overview of the current theoretical tools and experimental acceptance. I show that it is possible to discover the semi-elastic process with only 300 pb${^{-1}}$ of data but that the fully-elastic case is currently out of reach. I also illustrate how the use of forward proton tags can result in limits on the branching ratio for flavor changing neutral current decays of the top quark of the form $t\rightarrow u\gamma$ and $t\rightarrow c\gamma$ of $<0.39\cdot10^{-5}$ and $<0.97\cdot10^{-5}$, respectively, both of which would surpass the existing world limits by at least an order of magnitude.
\end{abstract}

\maketitle


\section{Introduction}
\label{sec:intro}

Top quarks are among the most well-studied and unique particles in the Standard Model (SM) of particle physics. Their high mass and short lifetime make them an ideal probe for understanding the nature of fundamental fermions. Since being discovered in 1995 by the CDF and DO collaborations at the Tevatron collider \cite{topdiscoveryCDF, topdiscoveryD0}, the production of top quarks by quark anti-quark annihilation, gluon-gluon fusion, and by quark-initiated electroweak interactions have been measured to impressive precision. However, the production of top quarks in proton-proton collisions where one or both of the protons remain intact, so-called semi-elastic or elastic production, respectively, remains undiscovered. In this letter, I outline the physics motivation for measuring such production modes, the theoretical tools and predictions that are currently available, and the unique way in which these processes can be used to search for signs of new physics. 

Elastic production occurs when one or both colliding particles (such as protons or ions) interact but do not dissociate and break up. These interactions are mediated either by pomerons ($\mathds{P}$) or photons ($\gamma$). Elastic production has been extensively studied in the context of total cross-sections of proton-proton and proton$-$anti-proton collisions and, more recently, in the context of the production of high-scale particles in interactions of the form $pp \rightarrow pXp$, where $X$ refers to a colorless, electrically-neutral final state such as $\ell^{+}\ell^{-}$. Such a process was observed in proton-proton collisions by the CMS collaboration \cite{cmslightbylight}, and the ATLAS experiment recently observed the $\mu^{+}\mu^{-}$ process in the collision of lead ions mediated by two photons \cite{atlaslightbylight}. Both the elastic and semi-elastic production of top quarks are considered in this letter.

When protons interact elastically they lose some energy and, as they leave the interaction point, they are swept out of the beam by magnets to maintain the integrity of the beam. Both the ATLAS and CMS experiments exploit this effect by placing dedicated ``forward proton'' detectors hundreds of meters down the beamline from the interaction point to detect and measure these deflected protons. These forward detectors can , therefore, be used to reliably identify elastic events. The AFP detector \cite{AFP} is used by the ATLAS detector to tag forward protons whereas the CT-PPS \cite{TOTEM} is used by the CMS detector. In this letter, I focus on the AFP and its potential to detect elastic collisions in tandem with the central ATLAS detector as well as the potential physics that may be extracted from observation of these processes.

\section{Production cross-section}

\begin{table*}
    \centering
    \begin{tabular}{l c c c c c}
    \toprule
    Generator Setting                         & \blank \pPttbar\ [pb]\blank & \blank \apttbar\ [pb]\blank & \blank \aPttbar\ [pb]\blank   & \blank \PPttbar\ [pb]\blank & \blank \aattbar\ [pb]\blank   \\
    \midrule
    SuperChic (isurv = 1)                     &              --             &             --              &               --              & $1.22(1)  \cdot 10^{-5}$    & $2.05(2)       \cdot 10^{-4}$ \\
    {\color{white}SuperChic }(isurv = 2)      &              --             &             --              &               --              & $3.21(2)  \cdot 10^{-5}$    & $2.06(1)       \cdot 10^{-4}$ \\
    {\color{white}SuperChic }(isurv = 3)      &              --             &             --              &               --              & $2.05(1)  \cdot 10^{-5}$    & $2.05(1)       \cdot 10^{-4}$ \\
    {\color{white}SuperChic }(isurv = 4)      &              --             &             --              &               --              & $1.59(1)  \cdot 10^{-5}$    & $2.06(1)       \cdot 10^{-4}$ \\
    {\color{white}SuperChic }(sfaci = false)  &              --             &             --              &               --              & $1.73(1)  \cdot 10^{-3}$    & $2.77(2)       \cdot 10^{-4}$ \\ 
    MadGraph                                  &              --             & $1.23$                      &               --              &            --               & $3.33\nil      \cdot 10^{-4}$ \\  
    PYTHIA (MPI: unchecked)                   &          $90.5(1)$          & $1.45$                      & $1.26(6) \cdot 10^{-1}$       &            --               & $4.56(2)       \cdot 10^{-4}$ \\
    {\color{white}PYTHIA} (MPI: checked)      &          $5.14(5)$          & $1.46$                      & $1.27(6) \cdot 10^{-1}$       &            --               & $4.57(2)       \cdot 10^{-4}$ \\
    FPMC\cite{marek}                          &              --             &             --              & $5.2\blank\nil \cdot 10^{-2}$ & $2.84\nil \cdot 10^{-2}$    & $3.4\nil\blank \cdot 10^{-4}$ \\
    \bottomrule
    \end{tabular}
    \caption{Summary of the various MC predictions for the elastic or semi-elastic production of $t\bar{t}$ pairs via pomerons or photons. A dash indicates that the given process is not implemented in the generator. The quoted uncertainties are purely statistical. When no uncertainties are quoted, the statistical uncertainty is negligible.}
    \label{tab:table2}
\end{table*}

There are several different types of elastic processes that are capable of producing top-quark pairs. They can broadly be grouped into QED and QCD processes, depending on if the particle mediating the interaction is a photon or a pomeron. For semi-elastic processes, I consider \pP\ and \gamp\, where the proton that emits the photon ($\gamma$) or pomeron ($\mathds{P}$) remains intact and the other dissociates. For fully elastic processes, I consider photon-photon fusion (\gamgam), double pomeron exchange (\PP), and photo-production (\aP). In all processes, the photon flux is modeled using the Equivalent Photon Approximation \cite{EPA} and a detailed discussion can be found in recent work~\cite{AAttbar2018}. Pomerons can be modeled non-perturbatively as a quasi-real particle with partonic structure, not unlike a proton, or perturbatively as a color-singlet exchange involving multiple gluon interactions (\emph{``the Durham model''}). For the semi-elastic cases considered here, the non-perturbative approach is followed whereas for double-pomeron exchange, the Durham model is used.

The \textsc{Pythia}~8.2~\cite{Pythia1,Pythia2} generator is capable of simulating all processes, with the exception of double pomeron exchange, and is used as the primary generator in the subsequent studies. The \textsc{MadGraph5}\_aMC@NLO~2.7.2 generator~\cite{madgraph} is capable of simulating events involving elastic photons but not pomerons. Is it also the only generator to be able to simulate new physics in an Effective Field Theory (EFT) format and is used for estimating the sensitivity of these processes to new physics in Section~\ref{sec:FCNC}. The SuperChic generator~\cite{superchic} is a dedicated central exclusive production generator and thus can only simulate fully elastic double pomeron or double photon exchange. Finally, the FPMC generator is capable of simulating all fully elastic processes but was not studied directly here. Instead, the predictions are taken from a previous study~\cite{marek}. All generators used in this study are interfaced to the NNPDF2.3 leading order PDF set~\cite{nnpdf}. Table~\ref{tab:table2} summarises the different predictions from each of these generators for their relevant processes. 

\begin{figure}
    \centering
    \includegraphics[width=0.475\textwidth]{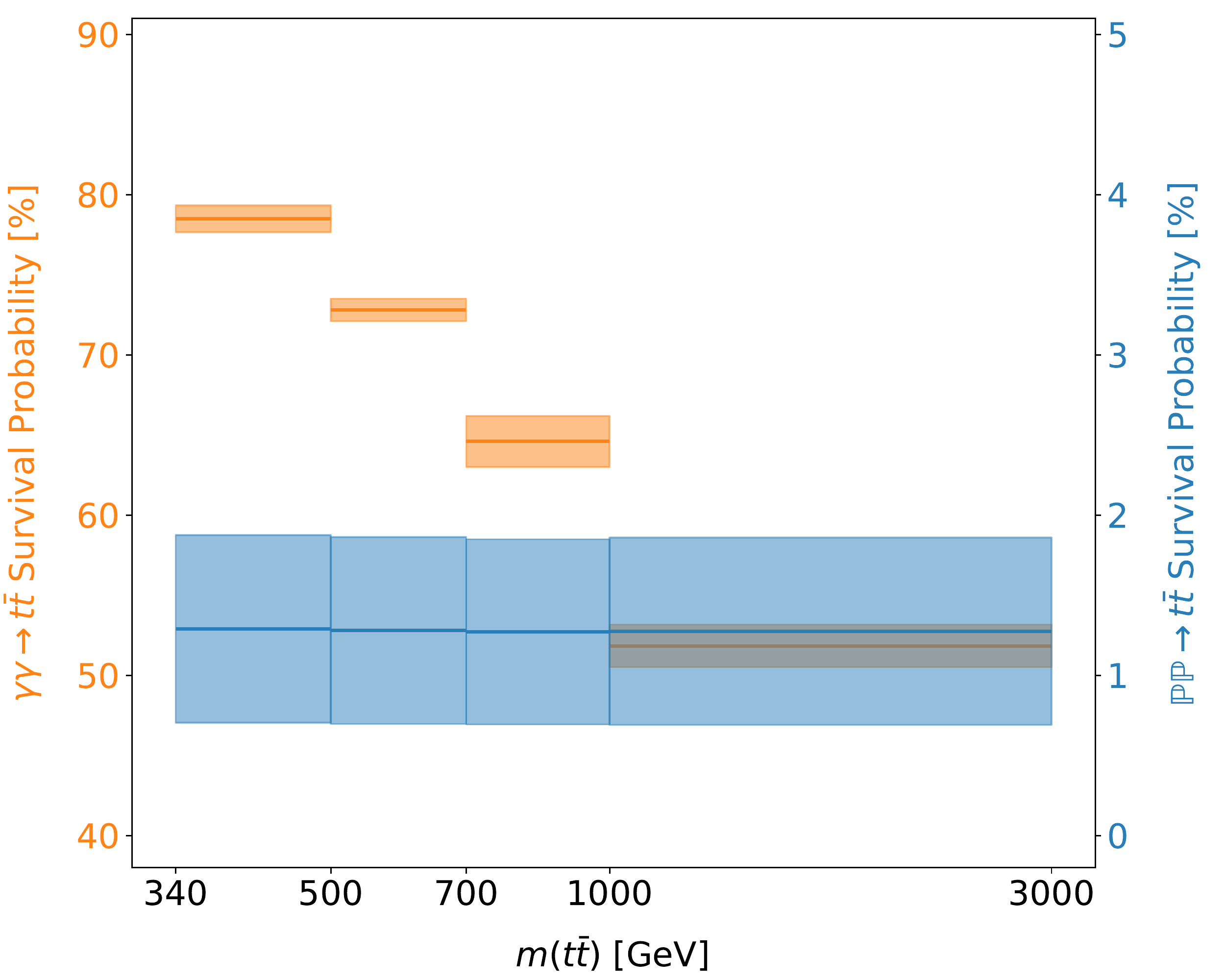}
    \caption{The survival probability for elastic top production, as determined by SuperChic~\cite{superchic}, as a function of the invariant mass of the $t\bar{t}$ system.}
    \label{fig:surv}
\end{figure}

A crucial consideration is the probability for the proton to remain intact in these interactions or, equivalently, the probability for further soft interactions to occur between the protons. This proton survival probability ($S_{eik}^2$) is proportional to the distance of the interactions and is therefore expected to be high for interactions involving photons, which favor long-distance interactions, and small for diffraction involving pomerons, which is a very short-distance interaction. The SuperChic generator uses a ``single-channel eikonal'' model to describe these additional soft interactions, where the proton is no longer considered to have partonic structure and all interactions are assumed to be identical and their rate described by Poisson statistics. For interactions involving pomerons, $S_{eik}^2$ is expected to be around $0.03$, whereas for photons it is expected to be relatively close to unity. Four different models for $S_{eik}^2$ are available in the SuperChic generator and the cross-sections for elastic processes with this generator are summarised in Table~\ref{tab:table2} (isruv$=1\rightarrow 4$) as well as with $S_{eik}$ to unity (sfaci=false). In Figure \ref{fig:surv}, the survival probability is estimated by taking the ratio between generated elastic $t\bar{t}$ events that either include or do not include the survival probability of the protons, as a function of the invariant mass of the $t\bar{t}$ system. For photon interactions, the survival probability is highest at low invariant mass at close to 80\% and drops to around 50\% at higher masses. In contrast, for pomeron interactions, the survival probability is extremely low but not dependent on the invariant mass of the system. The uncertainties on these curves are derived from variations of the survival model in SuperChic and have negligible statistical uncertainty. \textsc{Pythia} has the ability to check if multi-parton-interactions (MPI) have occurred between the protons, indicated in Table~\ref{tab:table2} with ``MPI: (un)checked'', and this can also be considered as a less-sophisticated estimate for $S^2_{eik}$. For these studies, the SuperChic approach for $S^2_{eik}$ is taken and only the ``MPI: unchecked'' values for the cross-section are used. 

As was alluded to earlier, each generator handles pomerons very differently, and therefore care should be taken when comparing the values in Table~\ref{tab:table2}. In the case of photon-initiated processes, however, each of the generators agrees relatively well, which is to be expected as they all use similar implementations of the EPA.

\section{Observing elastic events}

Given the production cross-sections described in Table~\ref{tab:table2}, the next pertinent question is how much data would be required to observe these processes? The cross-section for non-elastic $t\bar{t}$ production is between three and five orders of magnitude larger than any semi- or fully-elastic process and would be a primary background to any measurement. However, the inclusive process does not contain forward protons and by requiring the presence of at least one detected forward proton, this background can be largely eliminated if there are no spurious protons tags. A likely source for such tags comes from other elastic interactions occurring simultaneously, a common occurrence in the usual high pileup environment in which $t\bar{t}$ is studied. I, therefore, focus on special runs that are occasionally performed at the LHC with very low pileup where the average number of interactions per bunch crossing ($<\mu>$) is between 0.01 and 2 and in which spurious proton tags would be negligible. In the following sections I estimate the acceptance of the central and forward detectors of each of the elastic processes and calculate the amount of integrated luminosity needed to observe them using the following formula:
\begin{equation}
    \text{N}_{\text{events}} = \sigma \cdot \epsilon_{\text{BR}} \cdot A_{\text{central}} \cdot A_{\text{forward}} \cdot S_{eik}^{2}, 
\end{equation}
where ${N}_{\text{events}}$ is the number of expected events, $\sigma$ is the theoretical cross-section, defined in Table~\ref{tab:table2}, $\epsilon_{\text{BR}}$ is the branching ratio for the chosen decay mode of the $t\bar{t}$ final state, $A_{\text{central}}$ is the acceptance of the central ATLAS detector, $A_{\text{forward}}$ is the acceptance of the forward AFP detectors, and $S_{eik}^{2}$ is the proton survival probability. It should be noted that it is not expected that $S_{eik}^2$ factorizes for each elastic proton. Therefore the values of $S_{eik}^{2} = 0.8(0.03)$ are used for all photon(pomeron) processes, not just the fully-elastic cases from which they are derived.

\subsection{Forward detector acceptance}

\begin{figure*}
    \centering
    \includegraphics[width=0.475\textwidth]{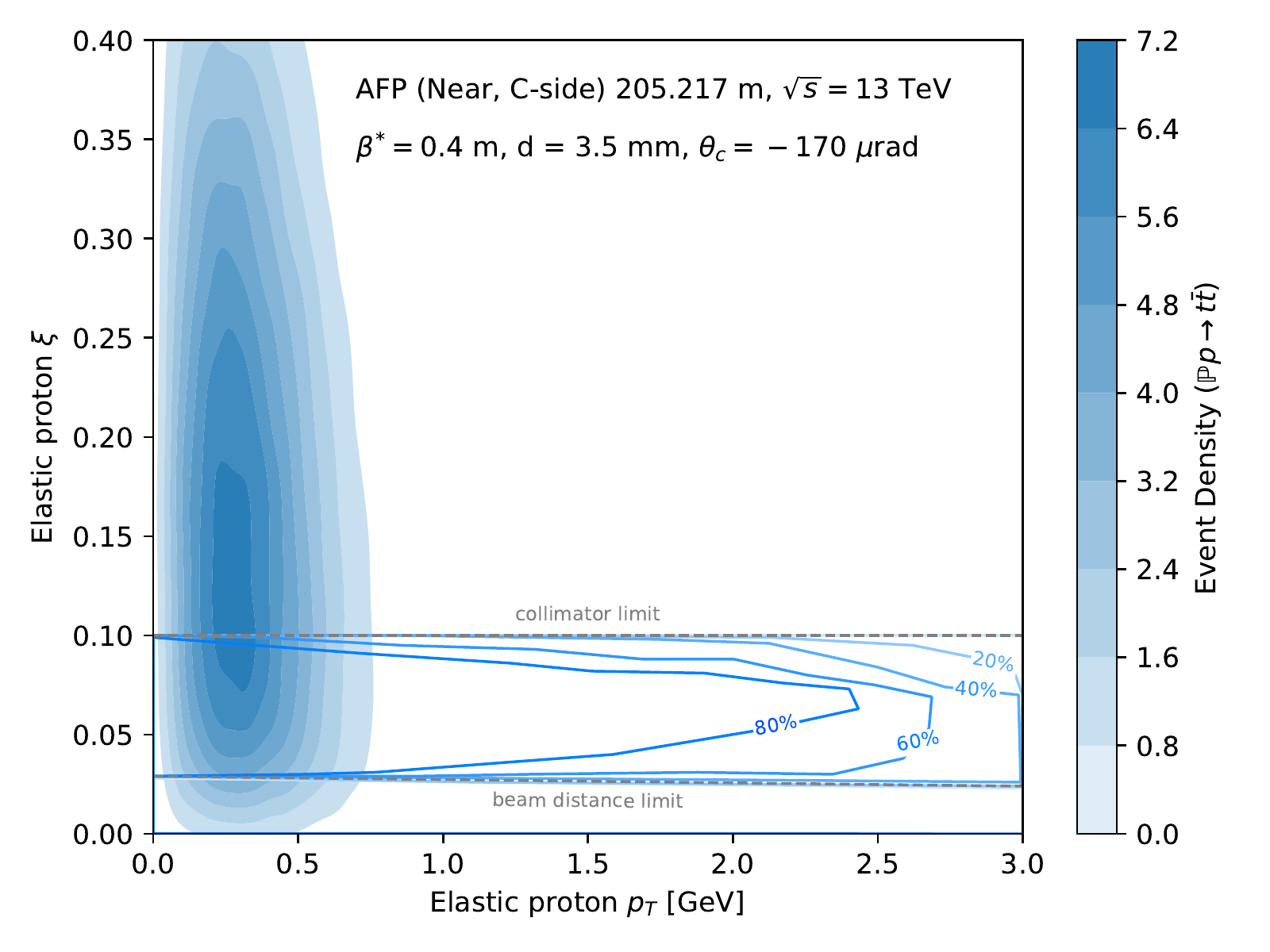}
    \includegraphics[width=0.475\textwidth]{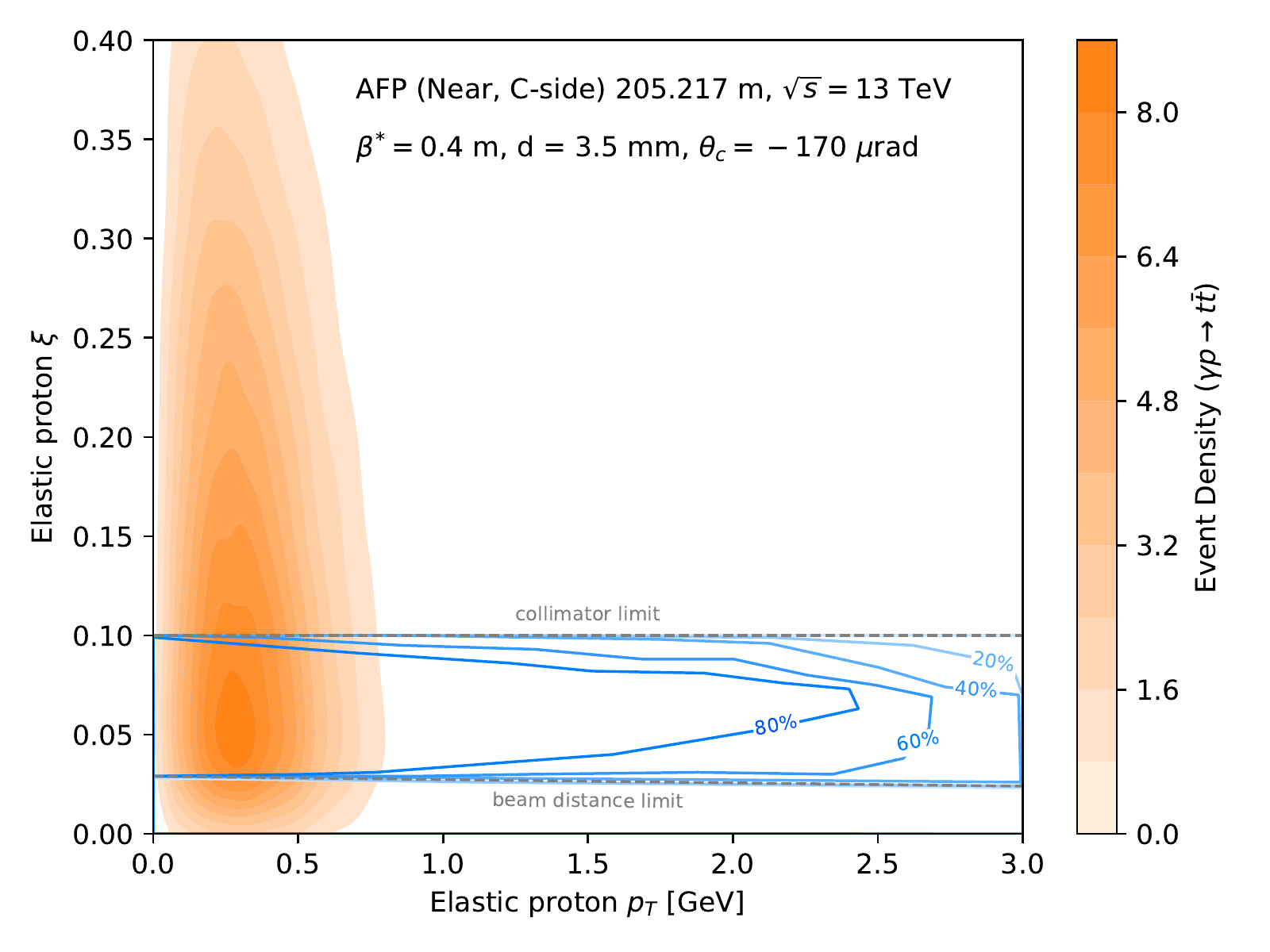}
    \caption{Density of events for elastic $t\bar{t}$ processes (pomeron initiated, left, photon initiated, right) as a function of the proton's $p_{T}$ and relative momentum loss ($\xi$). Acceptance contours for an ATLAS AFP station are overlaid for comparison.}
    \label{fig:afpprocesses}
\end{figure*}

The acceptance of the forward detectors is taken from publicly-available efficiency results from the AFP detectors as a function of the proton relative momentum loss ($\xi$) and proton $p_{T}$ \cite{Petousis:2673239}. These acceptances are strongly affected by factors such as; the distance of the AFP detectors from the beam, the settings of the collimators, and the beam optics. The efficiencies used in this study assume standard proton-proton beam conditions that were used during Run2, with a relatively small $\beta^*=0.4~\textrm{m}$ and that the distance of the detectors from the beam was $3.5~\textrm{mm}$. The result of these effects and the efficiency of the forward detectors themselves are illustrated for semi-elastic $\mathds{P}p \rightarrow t\bar{t}$ and $\gamma p \rightarrow t\bar{t}$ processes for a single AFP station on the elastic side of the event in Figure~\ref{fig:afpprocesses}. The acceptance loss below $\xi = 0.03$ is due to the distance of the detectors from the beam, whereas the efficiency loss above $\xi = 0.1$ is due to the settings of the ATLAS collimators. The protons from the processes considered in this study have $p_{T} < 1.4~\textrm{GeV}$, where the AFP detectors are highly efficient, and $0.0 < \xi < 0.8$, illustrated in Fig.~\ref{fig:afpprocesses}. The acceptance effects are therefore mostly sculpted by the $xi$ acceptance. The efficiencies for $\gamma$-induced processes are around 30\% and are around 20\% for $\mathds{P}$-induced processes. 

\subsection{Central detector acceptance}

Top quarks decay to a $b$-quark and $W$ boson in more than 99\% of cases. Top quark pair events are therefore categorized by the decay of the $W$ bosons as either dileptonic ($\epsilon_{BR}= 0.09$), semi-leptonic ($\epsilon_{BR}= 0.45$), or fully-hadronic ($\epsilon_{BR}= 0.44$), depending on the number of resultant charged leptons. All decay modes are considered in this study with the exception of decays containing taus. All of the elastic and semi-elastic processes are simulated using \textsc{Pythia} and particles are decayed and clustered into colorless hadrons until they are either stable or have a mean lifetime of longer than $10^{-6}$s. Hadrons are then clustered into jets using the Anti-kT algorithm~\cite{antikt} in the RIVET framework~\cite{rivet,rivet3}. All of the physics objects reconstructed in RIVET are identical to those used it ATLAS top analyses~\cite{atlastop}. Events are selected which contain at least one electron or muon and at least one $b$-tagged jet within the central detector acceptance. Charged leptons, such as electrons and muons, and hadronic jets have similar kinematic acceptances at both ATLAS and CMS. Here I take a simplified approach and simply require that all objects in the central detector satisfy:
\begin{equation}
    p_{T}(\ell, j) > 25~\mbox{GeV},~|\eta(\ell, j)| < 2.5
\end{equation}
It is possible that these requirements could be improved slightly due to the clean nature of low-pileup events. For example, the $p_{T}$ could be lower given favorable detector and trigger conditions, however, the effect of such changes is expected to be small. Leptons and jets are not detected and reconstructed with perfect efficiency and this must be taken into account. Here I use efficiencies from publicly-available object reconstruction, identification, and trigger efficiencies from the ATLAS experiment. Efficiencies are sometimes published for the full Run2 data (2015 - 2018) and sometimes by individual data-taking year. Efficiencies are chosen which best reflect low-$<\mu>$ data-taking conditions during 2017 when the AFP detectors were operating with high efficiency. Trigger and $b$-tagging efficiencies are parameterized as a function of the object's $p_{T}$. In cases where the object's $p_{T}$ is outside of the range provided from the ATLAS efficiency results, the nearest bin is used to derive the efficiency. Reconstruction and identification efficiencies are not strongly affected by the object kinematics and are taken as a single value. Several efficiencies are considered:
\begin{itemize}
    \item The trigger efficiencies of electrons and muons, assuming an unprescaled single lepton trigger with a low $p_{T}$ threshold.
    \item The reconstruction efficiency of muons.
    \item The identification efficiency of electrons.
    \item The $b$-tagging efficiency of jets containing $B$-hadrons.
\end{itemize}
Some efficiencies that are close to unity, such as jet identification efficiency, are not considered. The efficiency of ATLAS single-muon triggers with a threshold of at least 26 GeV for 2017 data is split between the barrel and endcap regions of the detector, and plateaus at 75\% and 85\%, respectively \cite{MuonTrigger}. The reconstruction efficiency of muons is taken to be 96\%\cite{MuonEfficiency}. The efficiency of ATLAS single-electron triggers with a threshold of at least 28 GeV for 2017 data varies between 75\% for threshold and up to 97\% for high $p_{T}$ \cite{ElectronTrigger}. The electron identification efficiency is taken to be 90\% based on low-$<\mu>$ ATLAS results \cite{ElectronID}. Finally, the efficiency for tagging jets containing $B$-hadrons is taken from the ATLAS MV1 algorithm with an $85\%$ efficiency working point and varies between $78\%$ and $89\%$, depending on the $p_{T}$ of the jet \cite{Btagging}. It should be noted that, in most cases, the efficiencies rise with lower pileup data. Since specific efficiencies are not publicly available for low-pileup runs for all objects, the ones used in this study should be conservative and are considerably lower than what is likely achievable by the experiments themselves. 

Dilepton events are selected by requiring exactly two leptons (electrons or muons) with opposite sign electric charge, and at least one $b$-tagged jet. Semi-leptonic events are selected similarly, except with a requirement of exactly one lepton. Since we are considering a low-pileup environment, and there are not many background elastic processes that can mimic this final-state, no requirement is made on the number of non-$b$-tagged jets. Finally, all hadronic events are selected by requiring exactly zero electrons or muons, exactly two $b$-tagged jets, and at least 4 non-$b$-tagged jets. This is a much tighter set of requirements than is used for the leptonic final states because there are many multi-jet elastic processes and thus a more stringent set of requirements is warranted. In practice, it may also be necessary to consider further requirements on the reconstructed top quark masses, similar to the method employed in recent ATLAS measurements~\cite{allhad}, but such techniques are not considered here. The dileptonic and semi-leptonic decay modes are considered in this study as these are relatively easy to trigger with high efficiency due to the presence of a charged lepton, but in the all hadronic case, we must rely on jet triggers. We assume a bespoke jet trigger could be created for a low-pileup environment with similar efficiency and performance to early 7 TeV data, with an efficiency of close to 100\% above 70 GeV. Therefore, for the all hadronic channel, at least one jet must have $p_{\rm{T}}$ greater than 70 GeV. These requirements result in a central detector efficiency of close to 2\% for the dilepton channel, 20\% for the semi-leptonic channel, and 5\% for the all hadronic channel, with little variation depending on the type of elastic process being considered.

\begin{table}
    \centering
    \begin{tabular}{c c c c}
    \toprule
       Process  & 100 pb$^{-1}$       & 300 pb$^{-1}$       & 1 fb$^{-1}$\\
    \midrule
        \gamgam & $9 \cdot 10^{-4}$    & $2.7 \cdot 10^{-3}$ & $9 \cdot 10^{-3}$ \\ 
        \PP     & $6 \cdot 10^{-5}$    & $1.7 \cdot 10^{-4}$ & $6 \cdot 10^{-4}$ \\
        \aP     & $1.6 \cdot 10^{-1}$  & $4.9 \cdot 10^{-1}$ & $1.6$             \\
        \gamp   & $9.4 \pm 0.3$        & $30 \pm 1$               & $94 \pm 3$        \\
        \pP     & $15 \pm 2$           & $40 \pm 7$               & $150 \pm 20$      \\
    \midrule
    Total       & $24 \pm 2$           & $70 \pm 7$               & $240 \pm 20$      \\
    \bottomrule
    \end{tabular}
    \label{tab:results}
    \caption{Number of expected events for elastic processes at various integrated luminosity scenarios. Statistical uncertainties are negligible. All results follow the rounding recommendations from the particle data group and totals are summed before rounding. When no uncertainty is quoted, the uncertainty was less significantly less than precision quoted.}
\end{table}

\subsection{Required Integrated Luminosity for Discovery}
\label{sec:results}

To observe a process, using purely statistical uncertainties, the standard criterion of an excess of 5 standard deviations from the null hypothesis can be met by observing 25 or more events above the background expectation (assuming only Poisson statistical errors). Three different benchmark delivered luminosities are considered at 13 TeV; 100~pb$^{-1}$, 300~pb$^{-1}$, and 1~fb$^{-1}$. The expected number of measured $t\bar{t}$ events for each of these benchmarks is presented in Table~\ref{tab:results}. Only statistical uncertainties from the cross-section calculation are considered and these are negligible. For fully elastic processes involving either one or two photons, the expected yields are well below one event and are therefore unlikely to be measurable in low-$\mu$ data. In contrast, the semi-elastic production could almost be measured with even the most pessimistic amount of low-$\mu$ data and should be observable (and perhaps even differentiated between pomeron- and photon-induced processes) with $300$ pb$^{-1}$ and above. It should be noted that the assumption of no background is generally true (given that statistical uncertainties on the data would be 10\% or higher at these expected number of events). The $t\bar{t}$ final state is not easily imitated by other SM signatures, and this is even more true for the elastic case. One process that would not form a relevant background but could form an additional signal is the associated production of a top quark and a $W$ boson, which can be produced semi-elastically, mediated by a photon, with roughly half the cross-section of the \gamp process. The central detector acceptance for this process would look very similar to the dileptonic and semi-leptonic cases for $t\bar{t}$ but would not pass the all hadronic selection (as there is only one $b$-tagged jet in the $tW$ final state). In the most optimistic luminosity case, the $tW$ process would add around 10 events to the total signal.

\begin{figure}
    \centering
    \includegraphics[width=0.32\textwidth]{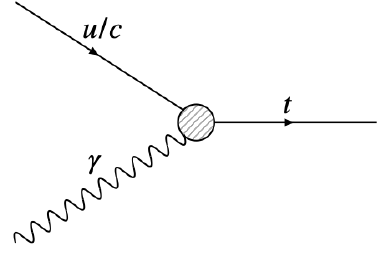}
    \caption{Feynman diagram illustrating photon-initiated processes with one or two top quarks in the final state and at least one FCNC coupling.}
    \label{fig:fcnc_diagrams}
\end{figure}

\section{Flavor Changing Neutral Currents}
\label{sec:FCNC}

Photon-initiated elastic processes are a potential laboratory for searching for the existence of flavor-changing neutral currents (FCNCs) of the form $t\rightarrow u/c \gamma$. The SM predicts that such currents can exist but that their existence is heavily suppressed. FCNCs could manifest in many elastic processes involving top quarks and photons but in most cases, there would be a significant SM background. One case, however, stands out as being uniquely sensitive. The production of a single top quark, with no associated quarks or bosons ($\gamma u \rightarrow t$), is something that effectively does not exist in the SM but could produce measurable numbers of events with relatively weak FCNCs. There is no other SM process that can imitate this signature, and an observation of it would be striking evidence for the presence of photon-mediated FCNC. This unique topology was already discussed in previous studies \cite{schannelFCNC2016}, however, the unique ability to suppress SM backgrounds by concentrating on the elastic process with a forward proton tag is discussed here for the first time. This process is modeled using \textsc{MadGraph5}\_aMC@NLO in an effective field theory context using the \emph{dim6top} model~\cite{dim6top}. This model allows 15 CP-conserving and 15 CP-violating degrees of freedom. We follow a similar EFT setup to the one used in \cite{ATLASFCNC}, with the added simplification that left-handed and right-handed couplings become degenerate in the $t\rightarrow u/c \gamma$ process and we, therefore, estimate limits on only two couplings:
\begin{equation}
    C_{uA}^{(3a)} \equiv C_{uA}^{(a3)} \equiv c_{W}C_{uB}^{(3a)} + s_{W}C_{uW}^{(3a)},
\end{equation}
where the index $a$ is 1 for up flavor quarks and 2 for charm flavor quarks. The SM predicts that the branching ratio for tops to decay to either an up quark or charm quark and a photon to be $4 \times 10^{-16}$ and $5 \times 10^{-14}$, respectively~\cite{snowmass}. The presence of many new physics models, such as a flavor violating two-Higgs-doublet-model (2HDM), can increase this considerably to $\mathcal{O}(10^{-7})$~\cite{2HDM}. The branching ratio for such couplings have already been probed by the ATLAS and CMS collaborations in top quark decays and have been constrained to the level of $ < \mathcal{O}(10^{-5})$ for $t\rightarrow \gamma u$ and $< \mathcal{O}(10^{-4})$ for $t\rightarrow \gamma c$ \cite{ATLASFCNC, CMSFCNC}. However, these analyses had to contend with huge SM cross-sections, relative to their potential FCNC signal strength, and must use complex neural networks to construct sensitive observables. Such experimental gymnastics are not necessary for elastic top production as the primary signature has no irreducible backgrounds and strong limits can be set based on a simple cut-and-count cross-section measurement. Though the study here explores the $\gamma p \rightarrow t$ process, the results are expressed as branching ratios for $t \rightarrow \gamma p$ to facilitate comparisons with existing limits from ATLAS and CMS. Using the same technique used to prototype the required amount of data to observe elastic processes in Section~\ref{sec:results} I extrapolate the limits that could be achieved by a lack of observation of the $\gamma \rightarrow t\bar{t}$ process with the three benchmark integrated luminosity values. Events are selected with exactly one $b$-tagged jet and one charged lepton ($e$/$\mu$), both with with $p_{T} > 25$ GeV and $|\eta| < 2.5$. The limits on the EFT couplings, and the corresponding branching ratios, are presented in Table~\ref{tab:limits}. 

In all scenarios, the BR$(t\rightarrow c\gamma)$ is improved compared to the best available world limits from ATLAS, by a factor of two in the most pessimistic case up to an order of magnitude in the optimistic scenario of one femtobarn of integrated luminosity. In the modest and optimistic scenarios, the limits for BR$(t\rightarrow u\gamma)$ are improved by a factor of 2 or 9, respectively. In the pessimistic case, limits are obtained that are a factor of two larger than the current best world limit. Each of these limits assumes that any background processes are perfectly modeled and subtracted from the data, a strong assumption. Nevertheless, they illustrate the potential of top quarks produced by elastic processes to provide competitive and possibly superior results than conventional searches.

\begin{table}
    \centering
    \begin{tabular}{c c c c c}
    \toprule
        Operator & 0.1 fb$^{-1}$ & 0.3 fb$^{-1}$ & 1.0 fb$^{-1}$ & ATLAS \cite{ATLASFCNC}\\
        \midrule
        $|C_{uW}^{(13)*} + C_{uB}^{(13)*}|$ & $<0.23$ & $<0.13$ & $<0.07$ & $<0.19$ \\
        $|C_{uW}^{(23)*} + C_{uB}^{(23)*}|$ & $<0.35$ & $<0.20$ & $<0.11$ & $<0.52$ \\
        \midrule
        BR$(t\rightarrow u\gamma)[10^{-5}]$ & $<4.05$  & $<1.35$  & $<0.39$  & $<2.8$   \\
        BR$(t\rightarrow c\gamma)[10^{-5}]$ & $<9.80$  & $<3.20$  & $<0.97$  & $<22$    \\       
        \bottomrule
    \end{tabular}
    \caption{Expected limits on EFT operators controlling left-handed interactions between top quarks and other up-type quarks as well as the most-recent limits from the ATLAS collaboration.}
    \label{tab:limits}
\end{table}

\section{Conclusion}

I have presented an example analysis for discovering the elastic production of top quarks at the LHC using forward proton tags, including an overview of the theoretical tools and experimental acceptance. I have shown that it is possible to discover the semi-elastic process with modest amounts of low pile-up data but that the fully elastic case is out of reach with low-$\mu$ data. Finally, I have illustrated how searches for FCNC involving top quarks and photons can be augmented and improved upon by using forward proton tags.

\begin{acknowledgments}
I wish to thank Lucian Harland-Lang and Ilkka Helenius from SuperChic and \textsc{Pythia}, respectively, for their invaluable help in understanding the behavior of their generators. I also wish to thank Andrew Pilkington for his advice and for first suggesting what has become an extremely fruitful avenue of research. I am supported by Royal Society fellowship grant  URF\textbackslash R1\textbackslash 191524. 
\end{acknowledgments}

\nocite{*}

\bibliography{apssamp}

\end{document}